\documentclass[%
 aip,
 amsmath,amssymb,
 reprint,%
]{revtex4-1}

\usepackage{graphicx}
\usepackage{dcolumn}
\usepackage{bm}

\usepackage[utf8]{inputenc}
\usepackage[T1]{fontenc}
\usepackage{etoolbox}
\usepackage{xcolor}
\usepackage{bm}

\usepackage{comment}

\newcommand{\mathbfh}[1]{\hat{\mathbf{#1}}}
\newcommand{\mi}{\mathrm{i}}

\makeatletter
\def\@email#1#2{%
 \endgroup
 \patchcmd{\titleblock@produce}
  {\frontmatter@RRAPformat}
  {\frontmatter@RRAPformat{\produce@RRAP{*#1\href{mailto:#2}{#2}}}\frontmatter@RRAPformat}
  {}{}
}%
\makeatother
\begin{document}

\title{Wading through the void: Exploring quantum friction and nonequilibrium fluctuations}
\author{D. Reiche}
\thanks{Now at Humboldt-Universität zu Berlin,
             Institut fur Physik,\\
             AG Optical Metrology \& Joint Lab Integrated Quantum Sensors, 12489 Berlin, Germany.}
\affiliation{Max-Born-Institut, 12489 Berlin, Germany}
\affiliation{Humboldt-Universität zu Berlin,
             Institut fur Physik,\\
             AG Theoretische Optik \& Photonik, 12489 Berlin, Germany}
\email{reiche@physik.hu-berlin.de}

\author{F. Intravaia}
\affiliation{Humboldt-Universität zu Berlin,
             Institut fur Physik,\\
             AG Theoretische Optik \& Photonik, 12489 Berlin, Germany}

\author{K. Busch}
\affiliation{Max-Born-Institut, 12489 Berlin, Germany}
\affiliation{Humboldt-Universität zu Berlin,
             Institut fur Physik,\\
             AG Theoretische Optik \& Photonik, 12489 Berlin, Germany}%

\date{\today}
\begin{abstract}
When two or more objects move relative to one another in vacuum, they experience a
drag force which, at zero temperature, usually goes under the name of quantum friction.
This contactless non-conservative interaction is mediated by the fluctuations of the
material-modified quantum electrodynamic vacuum and, hence, is purely quantum in nature.
Numerous investigations have revealed the richness of the mechanisms at work, thereby
stimulating novel theoretical and experimental approaches and identifying challenges
as well as opportunities.
In this article, we provide an overview of the physics surrounding quantum friction
and a perspective on recent developments.
\end{abstract}

\maketitle

\section{Introduction}
The interaction between an atom and a complex electromagnetic environment is one of
the oldest
~\cite{einstein10a,weisskopf30,purcell46}
and still one of the most frequently
~\cite{cronin09,hornberger12,keil16,lodahl17}
considered problems in modern physics. There are many reasons for that.

On the one hand, from the theoretical perspective, the problem lies at the confluence
of different physical disciplines
~\cite{mandel95,dalvit11}:
It starts with atomic physics and quantum electrodynamics and reaches via (nonequilibrium)
statistical physics and open quantum systems all the way to solid state physics.
As a result, studying the dynamics of an atom in the presence of macroscopic bodies
in various contexts represents an intriguing playground for testing the consistency and for advancing the predictive and interpretative quality of different physical theories.

On the other hand, from the experimental perspective, the subject of atom-field interactions
appears to be even more versatile. Nurtured by increasingly sophisticated designs
~\cite{reiserer15,amico17,gallego18,nayak19,meng20,aveline20},
the experimental control of (cold) atoms and atomic clouds has seen incredible advances since
the  pioneering work of the mid 1980s
~\cite{chu85,adams94}.
To name only a few examples, atom-field interactions are utilized for atomic clocks
and quantum sensing
~\cite{degen17},
atom interferometry
~\cite{schaff15,becker18}
on so-called atom chips
~\cite{keil16},
in diffraction experiments
~\cite{hornberger12},
and in modern information technology
~\cite{saffman10}.
Along these lines, the \emph{maturing quantum technolog[ies]}
~\cite{bongs19}
based on well-controlled atomic systems are
presently witnessing significant advances regarding
miniaturization with the goal of achieving portable high-precision sensors
~\cite{bongs19,frye21}.

\begin{figure}
  \begin{center}
    \includegraphics[width=0.45\textwidth]{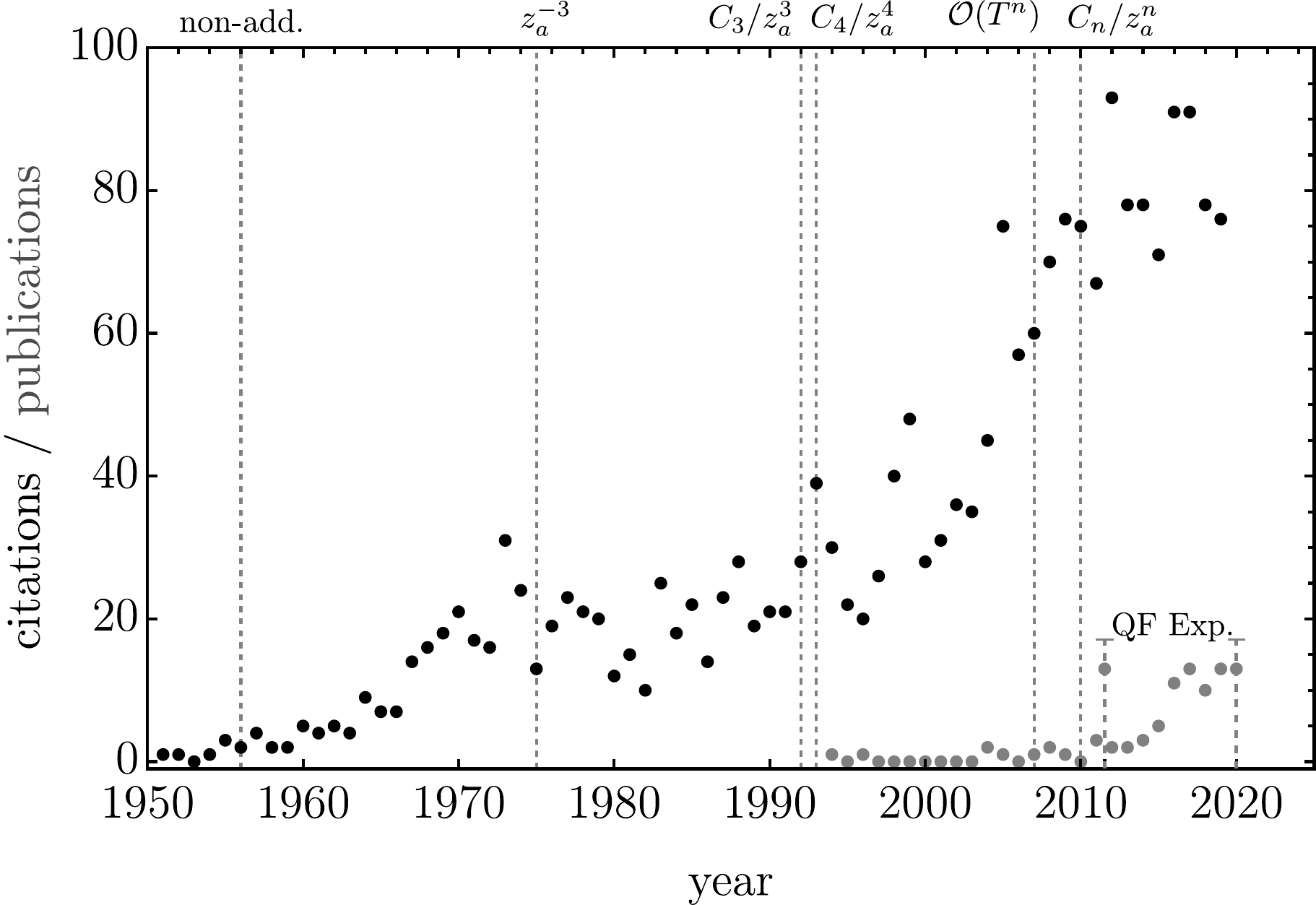}%
    \caption{
             Number of citations per year of Casimir and Polder's original
						 theoretical work
						 ~\cite{casimir48a} (black dots; inspired by similar plots in
						 Refs.~\cite{milonni94,lamoreaux05}).
             For comparison, in gray, the number of publications explicitly
						 mentioning \emph{quantum friction} is reported
						 \footnote{Data was collected using the ``Dimensions'' software
						          ~\cite{hook18}.
						           For the data on quantum friction, the search was restricted on
											 \emph{articles} in the area \emph{0202 Atomic, Molecular, Nuclear,
											 Particle and Plasma Physics} published in the journal family of
											 the American Physical Society. This reduces counting references
											 that use the phrase \emph{quantum friction} in a different context.
											 Since an explicit phrase was searched for, the list is not comprehensive
											 and, in particular, does not include older references such as Ref.
						           ~\cite{teodorovich76}.
                       }
					   The dashed lines denote certain important experimental milestones which
						 successively confirmed certain theoretical predictions (following Ref.
             ~\cite{dalvit11}): The force is
						 non-additive~\cite{derjaguin56}
						 (non-add.), scales as the inverse cubed atom surface distance $z_a^{-3}$
						 in the near-field~\cite{shih75}
						 and decreases even faster in the far-field ($z_a^{-4}$). The scaling in the
						 intermediate regime is more complicated, but generally follows a power law
						 ~\cite{bender10} ($z_a^{-n}$).
             The constants $C_{3,4,n}$ indicate that, additionally, a quantitative
						 analysis of the force's magnitude was performed
						 ~\cite{sandoghdar92,sukenik93,obrecht07}.
             A comprehensive study of the force's temperature dependence was conducted
						 in the late 2000s~\cite{obrecht07}
						 ($\mathcal{O}(T^n)$).
						 Lastly, two of the first proposals for measuring quantum friction are
						 denoted~\cite{volokitin11b,farias20}
						 (QF Exp.).
             \label{fig:CitationCP}}%
   \end{center}
\end{figure}

This miniaturization brings to the fore yet another intriguing aspect of atom-field interactions.
Indeed, one of the most surprising consequences of the quantum theory of light is that the
electromagnetic field never vanishes, despite the fact that it can be zero on average.
In close proximity to a material object, the electromagnetic local density of states is modified
by the corresponding boundary conditions that the field needs to obey at the material interface(s).
Atom(s) and molecules interacting with this field can, in turn, undergo a (complex) position-dependent shift in the energies characterizing their internal dynamics and in some limits also a change of their physicochemical properties ~\cite{cognee19,reiche20,kristensen20,kristensen21,garciavidal21}.
The corresponding effects are subsumed by the overarching term of `fluctuation-induced phenomena' since they can be attributed to the
fact that the interaction is mediated by the (material-modified) quantum and thermal fluctuations
of the field
~\cite{dalvit11}.
In particular, for an atom near an object, this interaction can usually lead to a modification
in its spontaneous emission rate, the so-called Purcell effect
~\cite{purcell46},
as well as an attractive force towards the material's interface, the so-called van der
Waals/Casimir-Polder force
~\cite{casimir48a}.
Both usually scale with power laws in the atom-interface separation and represent an important
element for experimental designs that aiming for closely spaced (commonly below some microns)
arrangements of their constituent elements.
It is thus not surprising that the importance of fluctuation-induced forces has recently
experienced a surge of interest alongside with experimental advancements
(see Fig. \ref{fig:CitationCP})
and is nowadays considered a standard feature of atomic trapping experiments
~\cite{druzhinina03,lin04,fortagh07,vetsch10,chan18}.

It is interesting to recall here that quantum-optical fluctuation-induced phenomena can also
be found beyond atomic systems. One closely related example is the counterpart of the
Casimir-Polder force, the so-called Casimir force which acts
between macroscopic bodies
~\cite{casimir48}.
Even though both effects were theoretically predicted at about the same time, they come
with rather distinct experimental challenges.
As alluded to before, probing the Casimir-Polder force has been greatly accelerated by
the spectacular advancements in the fields of atomic control and trapping in the late 1980s
~\cite{chu85,raab87}.
For the Casimir effect, for comparison, the main challenges lie in precisely
controlling the position of macroscopic objects with respect to each other at distances
below a micrometer.
Especially the latter has lead to the development of several experimental techniques
that only very recently have elevated the Casimir force to a well-established element
in the design and exploration of modern devices such as micro- and nano-electromechanical
systems (MEMS and NEMS)
~\cite{chan01a,rodriguez11,stange19}.

From the conceptual point of view, the above examples of fluctuation-induced forces
implicitly assume that the system is in global thermal equilibrium -- at least approximately
to leading order in the coupling between system and environment.
This assumption often leads to a significant simplification of the mathematical description
of the system. For instance, it allows for the application of some useful theorems
and closely related results of statistical physics.
In particular, one of them -- the fluctuation-dissipation theorem
~\cite{kubo66}
-- establishes a general connection between correlation functions and the corresponding
linear response functions of the system, both of which represent recurring quantities in
the evaluation of fluctuation-induced interactions.

However, most experiments rather frequently expose the system to nonequilibrium conditions. These include trapping laser fields, temperature gradients, or thermal and
electrical noise.
Since correlation functions are notoriously difficult to compute, the description of systems
out of equilibrium often has to rely on perturbative or approximate approaches.
They have proved quite successful when treating resonant phenomena or when dealing with
sufficiently small couplings between the system of interest and its environment
~\cite{carmichael93,mandel95}.
By the same token, the temptation to sapiently `stretch' the applicability range of equilibrium
techniques, such as the fluctuation-dissipation theorem, beyond global equilibrium is often
quite strong.
Indeed, focusing on the stationary situation, where a flow equilibrium of (constant) thermal
currents can be expected, it is often reasonable to assume that Thermal Equilibrium is
established Locally (LTE) for sufficiently well-separated subsystems
~\cite{polder71}.
Usually, this assumption works well for the case of thermal nonequilibrium, i.e. when different
bodies are held at different temperatures
~\cite{polder71,volokitin07,Biehs21},
allowing for the description of intriguing nonequilibrium effects
~\cite{dewilde06,thompson18,desutter19,Biehs21}
that are mostly connected to thermal fluctuations of the electromagnetic field.

The situation becomes considerably more subtle in \emph{mechanical nonequilibrium}, i.e.
when an atom or a nano-particle is in relative motion with respect to a complex
electromagnetic environment.
In terms of the system's dynamics, we can distinguish three different scenarios for the
trajectory of the particle's center-of-mass according to how Lorentz invariance is
broken:
(i) varying acceleration, (ii) constant acceleration, and (iii) zero acceleration in the
presence of other objects.
All three situations usually imply the existence of an external agent driving the motion.
Also, for simplicity, we consider the case where the system is (at least initially) at
zero temperature, thus putting quantum fluctuations in the spotlight.
Cases (i) and (ii) correspond, respectively, to the so-called dynamical Casimir effect
~\cite{dodonov20,nation12}
and the Fulling-Davis-deWitt-Unruh effect
~\cite{nation12}.
Both effects are often considered for a motion in absolute vacuum and typically focus is
laid on the motion-induced thermal-like excitation of the particles' internal degrees of
freedom and on the creation of real and entangled photons from the system's mechanical
energy.
While for these first two scenarios, (i) and (ii), the breaking of Lorentz invariance can
be directly connected to the nonzero acceleration, the situation is different for scenario
(iii) since it features the peculiarity of concentrating on constant velocity, i.e. zero
acceleration, and rather emphasizes the importance of the interaction with macroscopic
material objects.
In fact, due to dissipation and dispersion in these material objects (see Sec. \ref{Sec:ASQF}),
the material-modified electromagnetic fluctuations define a privileged reference frame with
respect to which the motion occurs.
As a consequence, the fluctuation-induced force acting on an object that moves parallel to
a material interface exhibits two components. The aforementioned component lateral to the
particle's motion that attracts the particle towards the interface and a novel component
parallel to the interface that decelerates the particle's motion.
At zero temperature, this second component of the force is called \emph{quantum friction}
~\cite{pendry97}
and will be the main object of interest in the present manuscript.

From the theoretical perspective -- due to the complex interplay of fluctuations,
nonequilibrium physics, material properties and backaction between system and
environment -- quantum friction has been shown recently to require special care in
its description with regard to the usage of equilibrium-based or perturbative
approaches.
Specifically, some of the approximations and the techniques discussed above produce
distinct scaling laws for the same phenomenon
~\cite{intravaia14,klatt17}
or simply fail in its quantitative description
~\cite{intravaia16a,intravaia19a,reiche20a}.

From the experimental perspective, quantum friction offers plenty of room for
investigations. In fact, until now, only a few proposals on how to test this effect
have appeared
~\cite{volokitin11b,volokitin16,farias20,oelschlaeger20,reiche21}
(see Fig. \ref{fig:CitationCP}).
This is particularly interesting since, to the best of our knowledge, the quantum
frictional force has so far eluded a direct and unambiguous experimental confirmation.

Although the general concept is valid for generic bodies in relative motion
~\cite{volokitin07,dedkov17},
quantum friction between an atom and a surface represents an excellent trade-off
between
(i) being a sufficiently approachable theoretical playground that contains all relevant
the elements characterizing a realistic physical system
and
(ii) being potentially relevant for future experimental setups and numerous applications.
Even more so, by closely inspecting Fig. \ref{fig:CitationCP}, one can even make a case
for mechanical nonequilibrium becoming one of the next workhorses of atom-surface
dispersion forces.
Next to an atom or an atom-like system (e.g. a nano-particle) moving near a macroscopic
body
~\cite{mahanty80,schaich80,tomassone97,kyasov01,intravaia14,intravaia15a,jentschura15,pieplow15,klatt16,lombardo17,rodriguezlopez18,viotti19,farias19}, the most common scenarios involve two planar macroscopic bodies
~\cite{teodorovich76,levitov89,polevoi90,pendry97,hoye11,maghrebi13,hoye19}.
A related, but slightly different setup consists of rotating particles in vacuum or
near other objects
~\cite{manjavacas10,zhao12,silveirinha14b,dedkov17a,stickler18,arita18,pan19}.

While in the remainder of the manuscript we focus on the zero temperature case, it
is important to recall that thermal fluctuations also give rise to and impact
the frictional dynamics of an object in motion.
For instance, for thermal fluctuations in the absence of material objects, it has
been shown that the motion in such a thermal vacuum induces a drag force that opposes
the motion ~\cite{mkrtchian03}. This behavior is strictly related to drag effects
that Einstein and Hopf have analyzed in their pioneering work
~\cite{einstein10a,milton20a}.
Eventually, both thermal and quantum effects have to be considered together when
characterizing the full dynamics of an object moving within a complex structured
electromagnetic environment at finite temperature
~\cite{dedkov02a,kyasov02,volokitin07,pieplow13,viotti19,Oelschlager21}.

Finally, it is worth noting that many aspects of quantum friction, including its
very existence~\cite{milton16}, have been actively debated in the recent past
~\cite{philbin09,volokitin11,philbin11}.
While the premise of the present perspective is that quantum friction is a real
physical phenomenon, we believe that the amount of discussions around this topic
reveals -- next to its lack of triviality -- also the richness of the
underlying physics.

\section{Atom-Surface Quantum Friction}\label{Sec:ASQF}

\begin{figure}
  \begin{center}
    \includegraphics[width=0.45\textwidth]{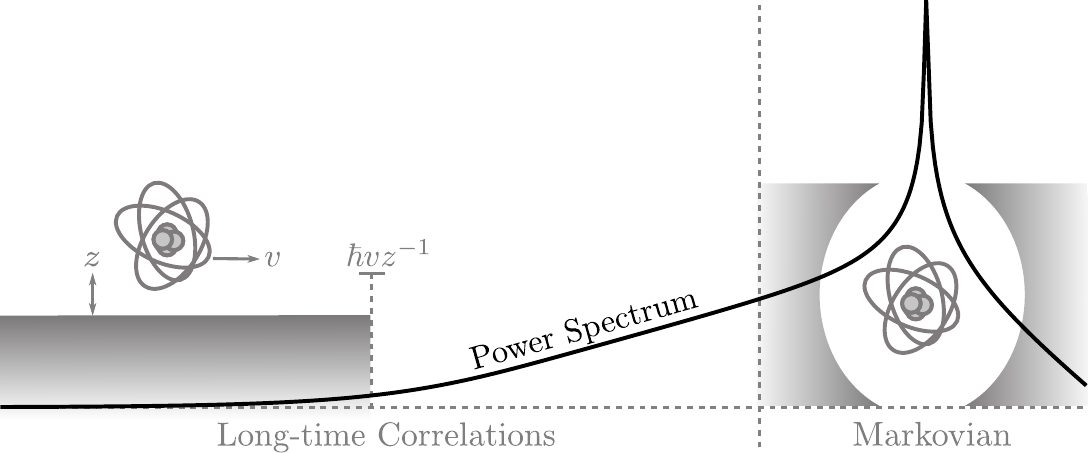}%
    \caption{Sketch of a typical power spectrum of the field fluctuations
		         $\underline{\chi}_v(\omega)$ (black curve) evaluated at Doppler-shifted
						 frequencies at later times, where the atom-field system can reach a
						 nonequilibrium steady-state.
						 We follow the derivations in Ref.
						~\cite{reiche20a}
						 and adopt the figure from Ref.
						~\cite{reiche21}.
             For simplicity, we work in the near-field of a planar conductor
						 which exhibits a single (surface plasmon-polariton) resonance.
             Near the resonance, Markovian methods are often well suited to
						 describe the situation as it is the case, e.g., in cavity quantum
						 electrodynamics.
             We depict the latter via the sketch that features the atom in
						 a small cavity.
For quantum friction, however, the relevant frequency scale $\hbar\omega\sim \hbar v/z_a$
lies in the low-frequency tail, where long-time correlations can lead to a strong non-Markovian
nature of the interaction (see Sec. \ref{Sec:ASQF}).
\label{fig:QuantumFriction}}%
\end{center}
\end{figure}

In order to obtain a first intuitive understanding of quantum friction, it is instructive
to consider a neutral atom moving in the vicinity of a macroscopic surface (see Fig.
\ref{fig:QuantumFriction}).
Ignoring, for the time being, possible trapping fields, every quantum operator describing
the individual constituents of the system is supposed to vanish in the mean. Still, due
to quantum fluctuations, the atom will possess an asymmetric charge distribution at a
certain instance of time $t$.
Neglecting magnetic effects and higher-order multipoles, we can describe the dynamics
of this charge asymmetry using the atom's electric dipole operator $\mathbfh{d}(t)$.
The field generated by this dipole obeys the laws of electromagnetism and the interaction
with the surface can be described in terms of total electric field operator
$\mathbfh{E}=\mathbfh{E}_0+\mathbfh{E}_I$, where $\mathbfh{E}_0$ denotes the field
in the absence of the dipole, which is generated by quantum fluctuations of charges and
currents in the material.
In turn, $\mathbfh{E}_I=\mathbfh{E}_I[\mathbfh{d}]$ represents the field that is induced
by the dipole. This can be approximately described in terms of an image dipole below the surface
~\cite{eberlein07,souza13}.
A self-consistent approach also requires that the field acts back onto the atomic dipole.
In the simplest case, the fluctuation-induced force $F$ can be derived from the (free)
energy $\langle\mathbfh{d}\cdot\mathbfh{E}\rangle\neq0$ and it can be roughly described
as a dipole-image-dipole interaction. When the atom is not moving, for symmetry reasons,
the image dipole is located exactly below the real dipole. This leads to an attractive
force normal to the surface.
When the atom is moving,  we can picture its image below the surface as being slightly
displaced because of the delay in the dynamics induced by the dissipative and the
dispersive features of the material. The dipole-image-dipole interaction is no longer
normal to the interface and the force then acquires the aforementioned component
parallel ($\parallel$) to the surface, i.e. the quantum frictional force $F$ which opposes the motion of the atom~\cite{intravaia11a}
\begin{align}
  \label{Eq:LorentzForce}
  F = \langle\mathbfh{d}\cdot\nabla_{\parallel}\mathbfh{E}\rangle
    = \langle\mathbfh{d}\cdot\nabla_{\parallel}\left(\mathbfh{E}_0
		                                              + \mathbfh{E}_I[\mathbfh{d}]
																						\right)\rangle.
\end{align}
Here, the angular brackets $\langle\cdot\rangle$ denote the respective quantum
average.
The density matrix relevant for the previous calculation depends on the particular scenario under investigation \cite{intravaia14,intravaia15a,klatt17} and we will later assume the system to be in a nonequilibrum steady-state (see Sec. \ref{Sec:IIB}).
Equation \eqref{Eq:LorentzForce} can be understood as the quantum equivalent  of the
Lorentz force acting on a dipole. Since the frictional force is, in general, not
conservative, it can not be defined in terms of a potential.

Further, Eq.~\eqref{Eq:LorentzForce} suggests that the dynamical light-matter
interaction is encoded in the electromagnetic properties of the environment as well
as in the correlator of the atomic dipole operator
$\langle\mathbfh{d}(t)\mathbfh{d}(t')\rangle$.
At late times, when dissipation in the system suppresses all transient behavior so that the friction balances the drive, the system approximately follows the center-of-mass
trajectory $\mathbf{r}_a\sim\mathbf{R}+\mathbf{v}t$ ($|\mathbf{v}|=v$ is the velocity
and $\mathbf{R}$ the position in the plane perpendicular to the direction of motion)
~\cite{intravaia14,hsiang18,reiche20a}.
The atomic dipole correlator can then be conveniently written in terms of the
(nonequilibrium) power-spectrum tensor
\begin{align}
  \label{Eq:PS}
  \underline{S}_v(\omega) = \lim_{t\to\infty} \int \mathrm{d}\tau~
	                          \langle \mathbfh{d}(t) \mathbfh{d}(t-\tau) \rangle
														\exp(\mi\omega \tau),
\end{align}
where the extra subscript $v$ indicates that the dipole is in motion.
We mention that, in equilibrium, the atomic power spectrum is connected to the
probability of emitting ($\omega>0$) or absorbing ($\omega<0$) a photon
~\cite{novotny12}.
This means that the atom is treated as an open quantum system that constantly
exchanges energy with its surrounding.
Mathematically speaking, the exchange of energy is manifest in the fact that
the atomic power spectrum is connected to the imaginary part of the coupled
(nonequilibrium) system's tensorial linear response function, which is directly
related to the dissipative features of the interaction
~\cite{dalvit11}.
In equilibrium, the exact relation between power spectrum and the dissipative
linear response is encoded in the celebrated fluctuation-dissipation theorem
~\cite{kubo66} and for our system it is closely related to the electromagnetic density of states
~\cite{joulain03}.
Without going into detail, we simply mention at this point that similar relations
can be found in certain nonequilibrium situations
~\cite{esposito09,seifert10,intravaia16a,reiche20a,hsiang20b}.
A general and easily applicable statement for arbitrary nonequilibrium processes
has, however, eluded the researchers so far.

The self-consistency of the previous description also allows us to make some general
considerations without specifying a concrete model for the system's nonequilibrium
dynamics.
Specifically, since also the dipole dynamics is influenced by the electromagnetic
environment, one has to expect that the power spectrum of the atomic dipole is
somehow non-perturbatively connected to the power spectrum of the material-modified
vacuum field
~\cite{reiche20a},
i.e.
\begin{align}
  \label{Eq:SelfConsistency}
  \underline{S}_v(\omega) = \underline{S}_v[\underline{\chi}_v(\omega)],
\end{align}
where
$ \underline{\chi}_v(\omega) = \lim_{t\to\infty} \int\mathrm{d}\tau
  \langle\mathbfh{E}_0(t) \mathbfh{E}_0(t-\tau) \rangle
	\exp(\mi\omega \tau)$
and the brackets in Eq.~\eqref{Eq:SelfConsistency} denote a model-dependent and usually non-trivial functional. The details and the
relevance of this relation can be used to justify or to assess the quality of
approaches based, for example, on the LTE or perturbative approximations.
This point proves to be quite important for the description of the
quantum frictional process, particularly when atoms are involved \cite{reiche20a}.

\subsection{On the scaling law of the quantum frictional interaction}
\label{sec1}

In order to gain further insights on the properties of quantum friction, it is useful
to assess the typical scaling laws of the force with respect to velocity $v$ and
atom-surface separation $z_a$.
To do so, it is instructive to first determine the relevant energy scales at work and consider situations commonly encountered in experiments.
Accordingly, we choose typical parameters which focus on specific nonrelativistic
regimes, thus to some extent foregoing full generality.
For simplicity, we continue to concentrate on the atom's electric dipole moment,
leaving higher-order multipole moments unattended. This implicitly places a lower
bound on the atom-surface separation for which our considerations hold. The closer
the atom gets to the surface, the more important higher-order multipole moments
can become
~\cite{raab05,crosse09}.
Further, we focus on the electric part of the frictional interaction only, since
the magnetic part is usually small, as long as typical experimentally accessible
parameters are used and common (non-magnetic) materials are involved.

Suppose, that the atom moves in vacuum at constant velocity $v\equiv|\mathbf{v}|$
and at constant distance $z_{a}$ parallel to a planar surface that separates space into two half-spaces, one of which is filled with a material.
Due to the atom's motion, its interaction with the electromagnetic magnetic field is
characterized by the Doppler-shifted frequency $\omega_{\mathrm{D}}=\omega-\mathbf{k}\cdot\mathbf{v} $.
Here, $\mathbf{k}$ denotes the component of the wavevector parallel to the surface.
The relevant values of $\mathbf{k}$ are determined by the atom-surface separation, i.e.  $|\mathbf{k}|\sim z_a^{-1}$.
Intuitively, this relation can be understood by considering a single plane wave traveling from the moving atom to the surface, becoming reflected and then reconnecting with the particle ~\cite{klatt16}.
Quite generally, this can be seen from a Weyl expansion
~\cite{li94a}
of the electromagnetic Green tensor that mediates the interaction and is connected
to the behavior of the electromagnetic density of states.
Since the frictional force must vanish for $v=0$, the impact of the motion on the
atom-surface interaction must thus be related to the Doppler-shift
\begin{align}
  \omega_{\mathrm{NEq}}\sim\mathbf{k}\cdot\mathbf{v}\sim v/z_a.
\end{align}
Note that this can, indeed, be confirmed by a more accurate calculation
~\cite{reiche20d}.
For separations $z_a\gtrsim1$ nm and velocities $v\lesssim3\times10^{-4}$c $\sim 90$ km/s,
the relevant energy scale evaluates to
\begin{align}\label{Eq:LowFreqEst}
  \hbar\omega_{\mathrm{NEq}}\sim\hbar\frac{v}{z_a}\lesssim 0.06\text{ eV}.
\end{align}
This simple estimate identifies the energy scale of the correlations in the electromagnetic field that are most relevant for quantum
friction. Comparing with the characteristic dipole resonance of, say, Rubidium ($\sim1.6$
eV for Rubidium-87
~\cite{steck})
or the surface-plasmon-polariton resonance of a conducting interface ($\sim 6.4$ eV
for gold
~\cite{barchiesi14}),
$\omega_{\mathrm{NEq}}$ appears to be rather small.
In other words, at reasonably small velocities, we can expect quantum friction to
be a \emph{low-frequency}, off-resonant phenomenon. With respect to the interaction
with the surface, this behavior puts the spotlight on the dissipative properties
of the material and, in general, on the low-frequency tails of the system's resonances.

Our analysis can be complemented with a rough estimate of the strength of the
corresponding light-matter interaction. For that, we again utilize our above-described
simple toy model of the atomic dipole interacting with its image dipole in the
material half-space.
Away from resonances, the induced field of the dipole approximately follows an
inverse-cubed dependence with distance
~\cite{joulain03},
i.e. $|\mathbf{E}_I|\propto |\mathbf{d}|/z_a^{3}$.
We note that a more precise argument follows from a consideration of the electromagnetic
density of states in the vicinity of the interface
~\cite{joulain03}.
When we further estimate the atom's dipole strength roughly through the elementary
charge $e$ and the first Bohr radius $a_0$, i.e. $|\mathbf{d}|\propto ea_0$, the
coupling rate between atom and material-modified electric field yields
$
|\mathbf{d}\mathbf{E}|/\hbar\sim (ea_0)^2/(\hbar \epsilon_0 z_a^3)
\approx 0.5
$ eV for a (rather optimistic) atom-surface separation $z_a$ of one nanometer.
Here, $\epsilon_0$ denotes is the vacuum permittivity and the square in the term
$(ea_0)^2$ underlines the fluctuation-induced character of the interaction that
originates from correlations.

These considerations show that, similar to other fluctuation-induced interactions,
quantum friction arises from small deviations with respect to the internal scales
that characterize the system's behavior. This opens several possible pathways for
the theoretical description of the phenomenon.
Specifically, for the evaluation of the dipole's power spectrum in Eq.~\eqref{Eq:PS}, we have
a plethora of theoretical approaches at our disposal, both perturbative
and not-perturbative
~\cite{carmichael93}.
Most of them agree in their qualitative as well as quantitative prediction close
to the system's resonances.
However, there might be substantial differences for the low-frequency tail of the
spectrum.
For instance, already in equilibrium at $v=0$, a description using the
Born-Markov approximation leads to a Lorentzian form for resonances in the power
spectrum.
This Lorentzian gives rise to a nearly constant and non-zero value for the
power spectrum at low frequencies and features the characteristic inverse quadratic
behavior at large frequencies, with a peak around the resonance
~\cite{intravaia16}.
For comparison, using the fluctuation-dissipation theorem, we would, close to the resonance, obtain neither a qualitative nor a relevant quantitative
difference to the Born-Markov approximation
~\cite{intravaia16}.
On the other hand, the actual low-frequency behavior is determined by (i) the
bosonic occupation number of the field and (ii) the local density of states at
low frequencies. In general, this can be very different from the aforementioned
behavior of the Lorentzian~\cite{intravaia14}.
When $v\neq0$, the situation is ever more complex and we refer to Refs.
~\cite{intravaia16,reiche20a,reiche21}
for a detailed discussion.

As discussed above, at low and moderate velocities low
frequencies are particularly relevant for quantum friction [see Eq. \eqref{Eq:LowFreqEst}], thereby exposing the differences between the various theoretical approaches.
Since both, frequency- and wave-vector scaling, are closely intertwined with the
velocity dependence (due to the Doppler shift) and the distance dependence (due
to the local density of states), the scientific literature abounds with different
predictions (see e.g. the overview provided in Ref.
~\cite{oelschlaeger20}).
Mostly, we find algebraic dependencies of the form $F\propto -v^{\delta}/z_a^{\kappa}$
with $\delta$ and $\kappa$ positive integers, but also non-algebraic forms have been
obtained
~\cite{reiche17,oelschlaeger18,reiche19,dedkov20a}.
Such rather distinct predictions are not necessarily mutually exclusive. The chosen
material models and how the calculations are carried out usually correspond to
different physical situations and one has to be very precise regarding their actual
range of applicability (see e.g. Ref.
~\cite{klatt21}).

For illustration, let us discuss an exactly solvable model for which the statistical
properties of the system's nonequilibrium steady state for the atom's dynamics can be
inferred in the linear regime
~\cite{intravaia19a}.
A single atom moving at nonrelativistic velocity $v$ parallel to a planar surface of
a half-space made from an Ohmic and spatially local material experiences the frictional
acceleration
\begin{align}\label{Eq:SimpleEstimateFriction}
  a
	\sim
  - \frac{18}{\pi^3}
    \frac{\hbar}{m}
    \alpha_0^2
    \rho^2
    \frac{v^3}{(2z_a)^{10}},
\end{align}
where $m$ and $\alpha_0$, respectively, denote the atom's mass and static polarizability.
Further, $\rho$ is connected to the dissipative low-frequency behavior of the local density
of states.
For conducting materials, $\rho$ corresponds to the specific resistance of the material.
We firstly note that the force is cubic in velocity and highly sensitive to the atom-surface
separation.
Indeed, as we have motivated above, these two issues are deeply rooted in the statistical
approach and the assumptions (self-consistency, Ohmic material, planar interface to a
half-space comprised of a spatially local material) that underlie the description. The
functional dependencies in Eq. \eqref{Eq:SimpleEstimateFriction} were also confirmed
by a careful perturbation theory
~\cite{intravaia15a,klatt17}.
Next, the force scales quadratically in $\alpha_0$. Since $\alpha_0\propto|\mathbf{d}|^2$,
this indicates that at least a fourth-order perturbation theory would be required to
capture this result. Indeed, it can be shown that the second-order perturbation theory
subtly cancels to zero when higher-order terms in the perturbation theory are considered
~\cite{intravaia14,intravaia16}.
Finally, the force is connected to the environment's dissipative features which are
characterized by $\rho$.
In particular, within the range of validity of Eq. \eqref{Eq:SimpleEstimateFriction},
a decrease of dissipation $\propto \rho$ leads to a quadratically decrease of the force.
This is in stark contrast to equilibrium dispersion forces, such as the Casimir and the
Casimir-Polder interactions, which especially at zero temperature are usually comparably
weakly affected by dissipation
~\cite{dalvit11}.

\subsection{The relevance of long-time correlations}
\label{Sec:IIB}
Given the number of theoretical predictions for exactly the same phenomenon (see the
previous section), the questions that remains is, \emph{what is the appropriate statistical
model for quantum friction?}
When attempting to answer this question, we vouch for thermodynamical arguments. In
order to elaborate on this point, we concentrate once more on the case of atomic motion
with constant velocity.

The moving atom and the nonequilibrium electromagnetic environment constitute a globally
closed system so that energy can only be redistributed among its constituents.
If we single out a subsystem, i.e. the atom, its dynamics is self-consistently connected
to the environmental degrees of freedom. Yet, dissipation in the system limits the
correlation time and introduces to the interaction the element of irreversibility.
When the atom is kept in constant motion and sufficient time has passed, different
energetic processes balance.
In this case, the dynamics can become stationary and the atomic degrees of freedom
reach a nonequilibrium steady-state
~\cite{barton16,hsiang18,lopez18}.
Sustaining this steady-state requires an external agent that drives the system and
exactly supplies the dissipated energy per unit time $P_{\mathrm{ext}}$ which compensates
the corresponding power lost due to the quantum frictional force, i.e.
$P_{\mathrm{ext}}=-vF$.
In this way, from the external perspective, a constant and non-vanishing energy current
originating from the drive is dissipated into the environment, mediated by the atomic
degrees of freedom
~\cite{sasa06,seifert10}.
In this nonequilibrium steady-state, a flow equilibrium is established so that the
average power entering the atomic subsystem $P_{\mathrm{in}}$ equals the average
power exiting $P_{\mathrm{out}}$, i.e.
\begin{align}
  \label{Eq:Thermodynamic}
  P_{\mathrm{in}}=P_{\mathrm{out}}.
\end{align}

The simplicity of the above relation should not hide the complexity that is required
for it to be fulfilled.
In fact, it has been shown that for non-Markovian systems with environmentally induced
dissipative degrees of freedom, such as atoms, the self-consistency condition [see
Eq. \eqref{Eq:SelfConsistency}] in combination with the long correlation times
$t_{\mathrm{NEq}}\sim|\omega_{\mathrm{NEq}}^{-1}|$ requires at least a fourth-order
perturbative treatment in the atomic dipole moment in order to satisfy
~\cite{reiche20a}
Eq. \eqref{Eq:Thermodynamic}.
Otherwise, one would neglect the energies corresponding to the long-time correlations
(for the above-discussed experimental scenarios
$t_{\mathrm{NEq}}\sim z_a/v$ is typically of the order of nanoseconds) and artificially
tilt the thermodynamic bookkeeping
~\cite{reiche20a}.
For comparison, if, instead of a self-consistent solution, we were to employ the
approximation of local thermal equilibrium (LTE), we would partially ignore these long-time correlations and obtain a power imbalance in the nonequilibrium
steady-state, i.e. $P_{\mathrm{in}}^{\mathrm{LTE}}-P_{\mathrm{out}}^{\mathrm{LTE}}>0$.
As a result, the atom would constantly heat up, thereby violating stability at late
times. The fascinating thing is that both theories, the self-consistent approach and
the LTE approximation, can lead to predictions with the same functional dependencies
on $v$ and $z_a$
~\cite{intravaia16a},
as long as the calculation is restricted to spatially local Ohmic materials
~\cite{reiche17,oelschlaeger18}.
Meanwhile, from the thermodynamic point of view, the differences and flaws
 of the different approaches are quite pronounced.

Again, this does not mean, however, that perturbative or approximate techniques are
unsuited for computing quantum frictional forces. Instead, in order to assess which
approach is more suitable, one has to carefully inspect the correlation times
and coupling strengths between the system's constituents.
For instance, if the system possesses sufficiently strongly coupled \emph{intrinsic}
degrees of freedom, local equilibration times could prevail against long-time
correlations that act on the global system. Specifically, when we replace the atom
with a polarizable nano-particle, the latter can feature a macroscopic number of
internal degrees of freedom that may act as a thermodynamic bath. For a metallic
nano-particle modeled, e.g., via a spatially local Drude material model, one obtains
a frictional acceleration
~\cite{intravaia16}
$a\propto -\alpha_0\gamma \rho v^3/z_a^7$, which can be sufficiently well described
within the LTE approximation (in this case, $\alpha_0$ is nano-particle's static
polarizability).
Comparing with Eq. \eqref{Eq:SimpleEstimateFriction}, we note that the deceleration
affecting nano-particles scales with the \emph{first} order of the particle's static
polarizability $\alpha_0$ and, hence, in second order of its dipole moment.
The internal correlation time scale, associated with the particle's intrinsic dissipation rate $\gamma$, reduces the required order in perturbation theory and
renders the equilibrium-based LTE approach a valid method.
This also holds for the thermodynamical perspective if the analysis of the
energetic bookkeeping is consistently performed at the same perturbative order
~\cite{hsiang18,reiche20a}.
Roughly, one can say that the LTE approximation works well for
$\gamma/\omega_a^2\gg \alpha_{0}\rho/z_{a}^{3}$,
where $\omega_a$ is, in this case, the dipolar resonance frequency of the nano-particle.
Similar arguments can be devised for the quantum friction between macroscopic
bodies
~\cite{dedkov17}
as well as for radiative heat transfer between two bodies
~\cite{polder71},
which explains the widespread success of the local thermal equilibrium assumption
in these fields.

Nonequilibrium dispersion forces derive from a complex interplay of different
length scales at work and come with many, often subtle and implicit assumptions.
Thermodynamic arguments provide an important and fundamental consistency
check for corresponding predictions.
Moreover, simple thermodynamic relations [e.g. Eq. \eqref{Eq:Thermodynamic}] also serve more concrete purposes such as being a benchmark for complex
numerical analysis
~\cite{reid17,molesky18,reiche20a}.

\subsection{Resonant friction, Cherenkov, and effective temperature}\label{Eq:ReonantFriction}

Equation \eqref{Eq:SimpleEstimateFriction} provides a reasonably good description of the
frictional force as long as $v/z_a$ is much smaller than any resonance in the system
~\cite{intravaia19a}.
For larger velocities, the kinetic energy thus infused into the system can become comparable
to the energies associated with resonances in the atom or the material which, in turn,
leads to another qualitative change in the force.
In fact, due to the then predominantly resonant interaction, Markovian methods are often, at least qualitatively, justified
~\cite{intravaia16,durnin21}
and the interaction can yield (sometimes non-algebraic) changes in scaling laws as well
as strong relative enhancements. This happens for atom-surface interactions
~\cite{intravaia16a,pieplow13}
and friction between macroscopic bodies
~\cite{guo14}
alike.

As pointed out in Sec. \ref{sec1}, it is noteworthy that for typical values that are
currently achieved in experiments (see Sec. \ref{Sec:laboratory}), the corresponding
energy scale $\hbar v/z_a$ is usually on the order of or smaller than tens of meV.
Since electro-optical resonances are typically of the order of tenths to a few of
electron volts, one would need to achieve extremely high (possibly relativistic)
particle velocities or very small particle-surface separations in order to exploit such resonant enhancement effects. This can be quite challenging, at least
for present-day experiments.
In order to alleviate these issues, some efforts have been undertaken to realize
resonant enhancement in more realistic scenarios using excited atomic states in
combination with more advanced geometric setups
~\cite{scheel09,durnin21}.

Intuitively, quantum friction can also be understood in terms of a re-distribution
(loss) of energy from the particle to its environment through all available channels.
This process is so general that the radiation alone can serve a similar purpose
without the need of any form of resistivity in the material.
As an illustration, we again consider the situation of a moving atom in the vicinity
of a planar surface. This time, however, the material is supposed to consist of a
perfectly lossless dielectric with constant and positive relative permittivity
$\epsilon = n_{\epsilon}^2>1$.
Then, one can show that frictional forces indeed still occur, but only for velocities
that exceed the speed of light in the medium, i.e. $v\geq c/n_{\epsilon}$.
Such a situation exhibits a strong connection to the Vavilov-Cherenkov effect
~\cite{cherenkov34}
and has already been studied in various settings
~\cite{maghrebi13,pieplow15,volokitin16b,intravaia16b}.
Moreover, the redistribution of energy and the associated relation between nonequilibrium
fluctuations and the dissipation in the system suggests that, in this nonequilibrium
steady-state, it becomes meaningful to study the (effective) thermalization properties
of the atom's intrinsic degrees of freedom.
In complete analogy to the Fulling-Davis-deWitt-Unruh effect, one can then define
an effective temperature perceived by the moving atom at \emph{constant velocity}
~\cite{intravaia16b,reiche21},
even though the system at large is held at zero temperature. The perceived effective
temperature is solely induced by the breaking of Lorentz invariance by means of the
atom's motion in proximity to the material interface.


\section{The overlooked and the unintuitive}\label{Sec:III}

The discussion on the statistical modeling of the low frequency regime of atom-surface
friction and its connection to the broader scope of nonequilibrium physics goes way
beyond an academic discussion of finding the appropriate technique for carrying out
the calculations.
In fact, exploring the depths of the advancements in theoretical descriptions leads
to fascinating ``flowers that blossom off the main road''.
In recent years, the peculiar relevance of unusual length scales in quantum friction
has instigated a number of refined investigations that have brought to the surface
previously overlooked effects. Without being exhaustive, we provide in the following
a small bouquet.

It is probably most natural to compare the predictions for quantum friction to some
of the characteristics of classical friction. For one, quantum friction is contactless,
a pure quantum effect and features an anomalous power-law dependence on velocity,
especially when compared to the classical Stokes friction that prevails at low
velocities and Reynolds numbers [see Eq. \eqref{Eq:SimpleEstimateFriction}].
That differences aside, there is more than meets the eye.
The comparison with the classical case could also remind us that there are
different kinds of friction: sliding and rolling friction, the latter implying a
specific interplay of the translational and of the rotational degrees of freedom
of the system.
At first sight, for an atom, this distinction might appear strange if not superfluous. Still, the comparison can become more clear if one explores if
and how angular momentum is transferred from the electromagnetic environment to the
atom during its motion. It turns out that, if the moving atom can exchange angular
momentum with radiation, the frictional force acquires a corresponding contribution,
$F^r$, that tends to diminish the `naive' result $F^t$ which does not take into account
the angular momentum transfer
~\cite{intravaia19a}.
For instance, within the self-consistent description considered above, in the situation
of an atom moving parallel to a planar surface, one obtains
\begin{align}\label{Eq:RotFric}
  F^r/F^t\approx -5/7,
\end{align}
where the negative sign indicates the reduction of the overall frictional force $F=F^t+F^r$.
Moreover, by defining the mean angular momentum tensor
$\bm{L} \propto \langle\hat{\mathbf{d}} \times \dot{\hat{\mathbf{d}}}\rangle$, we can
define an effective ``rotation frequency'' $\Omega$ (in 3D) of the average atomic dipole
moment
~\cite{intravaia19a},
very much resembling the situation of classical rolling friction.
However, in the quantum case, the \emph{sense} of rotation is completely off.
Due to the complex interplay of the nonequilibrium dynamics with phenomena such as the anomalous Doppler effect
~\cite{nezlin76}
and spin-momentum locking of light
~\cite{bliokh15,lodahl17},
the interaction between the atom and the electromagnetic excitations in the vicinity
of the material surface gives rise to a net absorption of angular momentum by the atom.
Intriguingly, the angular momentum carries a sign which is opposite to that of
its classical equivalent. One can say that the particle ``rotates'' in a (classically)
counterintuitive sense.
Specifically, if the atom moves along an axis from left to right, the sense of rotation
would be \emph{counter}-clockwise, i.e. exactly opposite to the classical counterpart of
a wheel rolling in the same direction over the surface
~\cite{intravaia19a,reiche20d}.
Surprisingly, this rotation is solely induced by the lateral motion of the atom and
is thus to be distinguished from another interesting situation where a rotation
of a (nano-)particle is enforced externally
~\cite{rodriguez-fortuno15,manjavacas17}.
This latter scenario, leads to quantum frictional torque
~\cite{zhao12}
and a lateral force on nano-particles close to a surface
~\cite{manjavacas17}
as well as to the transfer of angular momentum along a chain of rotating nano-particles
~\cite{sanders19}.

Another interesting aspect of the frictional interaction is its strong dependence on the field's local
density of states at $\omega \to 0$ (see Eq.
\eqref{Eq:SimpleEstimateFriction} and also the discussion below).
We note that the parameter $\rho$ enters quadratically into the evaluation of the force.
In consequence, we can expect the friction to be quite sensitive to
(i) the nature of the available dissipative channels
and
(ii) the spatial distribution of the corresponding material.
In simple terms, the more dissipation, the stronger the frictional force.
On the one hand, this points towards a judicious material choice in order to modify or control the interaction. For instance, semi-conductors such as Germanium
or n-doped silicon often feature stronger intrinsic low-frequency lattice-related
damping mechanisms than conductors. This can be proven to directly show in the
friction's strength
~\cite{oelschlaeger18}.
Moreover, one can also think of using layered structures or similarly manufactured
complex crystals, specifically hyperbolic metamaterials, in order to tailor the
dissipative profile of the material via the density of states
~\cite{oelschlaeger18}.
On the other hand, even for comparably simple materials such as ordinary and
homogeneous conductors, additional damping mechanisms can contribute to the
interaction. For instance, when the separation between the moving objects in
relative motion approaches the order of the electron's mean free path in the
conductor ($\sim40$ nm for gold
~\cite{gall16}),
the electromagnetic fluctuations that dominate the interaction can resolve
the ballistic motion of the material's electrons (more precisely, quasi-particle
excitations
~\cite{pines66})
which can lead to a coherent energy transfer between radiation and particles.
This process is independent of electronic collisions at impurities, phonons
or other electrons and is commonly referred to as Landau-damping
~\cite{landau46}.
In the case of quantum friction, Landau damping has been shown to lead to a
significant enhancement of the force which can reach several orders of magnitude
~\cite{volokitin03a,volokitin07,despoja11,reiche17,reiche19,dedkov20a}.
From the broader perspective, Landau damping can be one consequence of the
spatially nonlocal description of the material (although not every nonlocal
description necessarily includes Landau damping
~\cite{reiche20}).
In the context of quantum friction, spatial nonlocality has also been
investigated in configurations involving graphene or other topological
materials
~\cite{volokitin11b,morgado17,farias18}.
It is interesting to notice that, for equilibrium fluctuation-induced forces,
spatially nonlocal material models are often ambivalent.
Due to the interaction's broad-band nature and its usually less pronounced
connection to dissipation, the strength of the force at equilibrium is often
sufficiently well described by spatially local material models, e.g. the
Drude-Lorentz model
~\cite{dalvit11}
and corrections due to nonlocality are small, especially at low temperatures or short separations -- at least in simple geometries and for common
materials
~\cite{intravaia15}.
When more exotic materials such as graphene are involved
~\cite{klimchitskaya15a,rodriguezlopez17,egerland19}
or when thermodynamical aspects are considered
~\cite{svetovoy05,svetovoy08,reiche20},
the role played by nonlocality can be rather relevant.

With regards to the geometrical setting, the quadratic scaling of the frictional
force with the local density of states, outlined in Eq. \eqref{Eq:SimpleEstimateFriction}
by the behavior as a function of $\rho$, suggests a strongly non-additive behavior.
In other words, two times the number of interacting bodies surrounding the atom
does not necessarily give twice the frictional force, but can yield a very much
different result.
For instance, if $F$ is the force experienced by an atom that moves close to the
surface of an object and along its translationally invariant direction, the
force $\tilde{F}$ along the same axis of invariance of an ensemble of $N$ bodies
can reach the value
\begin{align}
  \tilde{F}\sim \phi N^2 F.
\end{align}
Here, $\phi>1$ is a prefactor related to the contribution to the force
due to the exchange of angular momentum between the atom and the radiation
[see Eq. \eqref{Eq:SimpleEstimateFriction}].
If we choose a rotationally symmetric configuration around the direction of motion,
the net transfer of angular momentum can be suppressed, and estimates based on
a planar interface
~\cite{intravaia19a}
gives $\phi\sim 3.5$. For instance, considering an atom moving at the center
of a planar cavity ($N=2$) or a waveguide ($N\sim 4$), we would expect a factor of $\sim 7$ or $\sim 14$ as a non-additive ``correction'',
respectively. With respect to the single planar interface, this means that
instead of a factor of $2$ and $4$, the force should be, respectively,
$\sim14$ and $\sim56$ times stronger.
More careful investigations have shown that, indeed, enhancements between one
or two orders of magnitude can be attained
~\cite{reiche20d}.
For comparison, although non-additive effects with respect to geometry are well-known
in the context of equilibrium Casimir(-Polder) forces, the corresponding non-additive
corrections are typically smaller, around the order of the unity or less, leading to enhancements that are usually smaller than a factor of two~\cite{rodriguez11}.
In this context, it is interesting to note for future investigations that, given
$F\propto\alpha_0^2$ in Eq. \eqref{Eq:SimpleEstimateFriction}, a similar behavior
can be expected to originate from the particle's polarizability.

Taken together, both the material properties themselves and the material's spatial
distribution carry considerable potential for controlling and tailoring the quantum
frictional force.
It might thus be no surprise that ensuing calculations in more advanced geometries
have been conducted just recently
~\cite{dedkov21}
and numerical optimizations of experimentally relevant parameters in the spirit
of emerging design- and inverse-design efforts
~\cite{molesky18,bennett20}
appear to be most promising for future work.


\section{In the laboratory}\label{Sec:laboratory}

The best physical theory has little relevance without its experimental reality.
Quantum friction is certainly no exception, specifically because, to the best
of our knowledge, it has thus far eluded direct experimental confirmation.
Probably, one of the major issues in this context is that, in the most common
situations, the effect is weaker than the sensitivity of the majority of
present-day experimental instruments.
To some extent, this is reminiscent of the circumstances
shortly after the prediction of London dispersion forces at the beginning of
the 1930s and of the Casimir force/ Casimir-Polder interaction at the end of
the 1940s. The first convincing measurements of these phenomena were
possible only some decades later.

Concerning the frictional interaction, if we consider, for instance, a Rubidium-87
atom moving parallel to a planar interface made from a gold layer much thicker
than the atom-surface separation, our simple estimate in Eq.~\eqref{Eq:SimpleEstimateFriction}
yields the frictional acceleration
\begin{align}
  \label{Eq:NumEstimateAcc}
  a\approx-3
  ~
  \left[\frac{v}{v_s}\right]^3
  \left[
  \frac{z_a}{\text{nm}}
  \right]^{-10}   \frac{\mu\text{m}}{\text{s}^2}.
\end{align}
Here, we have chosen an atom-surface separation $z_a$ of one nanometer and the
velocity of the speed of sound in air $v_s \approx 340$m/s as reference scales.
Rubidium and gold were chosen due to their widespread use in many experiments.
If one was to realize such conditions, the estimated acceleration would indeed be
comparable to the sensitivity of modern atom interferometric setups as they can routinely
achieve a precision below a micrometer per square seconds
~\cite{abend20a}.
A more ``experimentally friendly'' situation with an atom at a distance of a few
hundred nanometers (or even micrometers), confronts us, however, with a force
that is decreased by \emph{twenty to thirty} orders of magnitude.
From Eq.~\eqref{Eq:NumEstimateAcc}, a possible strategy to compensate for this
decrease and to enhance the force would be to increase the velocity, keeping
it within the non-relativistic limit.
Velocities more than an order of magnitude larger than $v_{s}$ have been achieved
in interferometric experiments with atom beams impinging on diffraction
gratings
~\cite{perreault05,cronin09,hornberger12,brand19}
(see also below).
Alternatively one could use a suitable combination of atomic species and substrate
material that, respectively, features an increased polarizability over mass ratio
and a larger value for $\rho$.
Specifically, Eq.~\eqref{Eq:NumEstimateAcc} [as well as Eq. \eqref{Eq:SimpleEstimateFriction}] reveals that good conductors, despite
being widely used in atom manipulation and trapping technologies, are not suited
for a detection of quantum friction.
For instance, using (doped) semiconductors instead of gold can yield an eight to
twelve orders of magnitude stronger frictional interaction ~\cite{oelschlaeger20}. In this case, also
nonlocal effects and Landau damping can be relevant and offer additional dissipation
channels, eventually increasing the frictional interaction even further.
Working with semiconductors can require, however, a level of modeling which is more
involved than for simple metals.
By the same token, working with Lithium instead of Rubidium gives roughly half the
polarizability at only one tenth of the mass, thus enhancing the frictional acceleration
by roughly a factor of three.
Such possibilities represent interesting pathways to optimize the design of experiments
that aim at measuring quantum friction (see Ref.
~\cite{oelschlaeger20}
for a comprehensive review).
Still, compensating for the force's immense decrease in strength associated with
larger atom-surface separations remains nonetheless challenging. Consequently, some
authors have questioned the feasibility of ever measuring quantum friction
~\cite{schaich80,milton16}.

For context, it it necessary to review some currently available experimental protocols
that are specifically targeting atom-surface interactions and might be considered
for the detection of quantum friction.  Without claiming to be anywhere close to
an exhaustive list, we may identify the following approaches:
Atom interferometry using either cold atoms
~\cite{abend20a}
or diffraction at material gratings
~\cite{cronin09,hornberger12},
and detection via nitrogen-vacancy centers
~\cite{schirhagl14}
instead of atoms. They all come with their different strengths and drawbacks.

For instance, cold atom interferometers are extremely sensitive
~\cite{abend20a}
(accelerations $\lesssim \mu$m/s$^2$ can be measured) and feature long interaction
times (of the order of milliseconds to seconds).
However, they are (typically) characterized by rather large atom-surface separations
(hundreds of nanometers and larger) and rather small velocities (centimeters to
meters per second or slower).
The speed and distance issue is directly addressed by another form of interferometry
where atomic beams are diffracted at a material grating. In this atom interferometer,
the interaction between the atoms and the slits of the grating leads to a characteristic
interference pattern on the detection screen. Typical setups allow for achieving very
high velocities (of the order of $\sim$km/s or higher
~\cite{brand19})
and narrow slit widths of the order of a few tens of nanometers naturally result in
rather small atom-surface separations. However, in order to achieve a high-resolution
interference pattern, the gratings are typically quite thin (hundred nanometers and
thinner) which leads to a short interaction time. In addition, uncertainties in the
geometrical structure and in other sources of interaction (e.g. stray fields
~\cite{cronin09,hornberger12})
may reduce the accuracy of the measurement.
Lastly, it can be convenient to consider atom-like systems instead of atoms. Prominent
examples are color-defects in nano-diamonds
~\cite{aharonovich11}
such as the nitrogen-vacancy (NV)
center
~\cite{doherty13}.
One of their practical advantages is that the nano-diamond embedding the defect
allows for an exquisite position control.
For example, glued to the tip of an atomic force microscope, a NV-center in a
nano-diamond can be controllably brought to a distance of just a few nanometers from
a flat surface.
This capacity of control has very recently lead to an experimental proposal for
detecting the effect of quantum friction on the internal dynamics of a NV-center
~\cite{farias20}:
The authors have shown that a separation of a few nanometers can be upheld
when the surface (a disk) is rotating with high frequency, giving rise to relative
velocities up to a few km/s.
At this point, it also is opportune to note that experiments measuring the interaction between the tip of an oscillating or fluctuating cantilever in an atomic force microscope or in a tuning fork and a flat surface have reported the existence of a contactless separation-dependent friction ~\cite{dorofeyev99,karrai00,gotsmann01,stipe01,kuehn06}.
Different materials have been used for both the tip (silicon, gold or gold-coated) and the surface (graphite, a polymer film above a gold layer, gold-coated mica).
However, the measured drag force features a different functional dependence on the
distance from the surface and a strength which is several orders of magnitude larger
than those predicted by Eq.~\eqref{Eq:SimpleEstimateFriction} and other studies
~\cite{chumak04,volokitin05,kuehn06b}.
A number of possible reasons for this behavior have been considered
~\cite{dorofeyev01,zurita-zanchez04}.
They include the presence of stray fields due to the polycrystalline structure of
the surface and adsorbates deposited on it
~\cite{stipe01,volokitin05}.
Nonetheless, a consensus on a definitive explanation of this striking result is still missing.
Still, these experiments have inspired proposals to measure quantum friction
based on the idea of keeping a cantilever at a fixed position above a silicon dioxide
substrate that is covered with a graphene layer through which an electric current is
running
~\cite{volokitin11b,volokitin16,shapiro17}.

In any of the aforementioned cases, the task of measuring the value predicted
in Eq.~\eqref{Eq:NumEstimateAcc} remains challenging and definitely requires the
optimization and the perfectioning of the experimental protocols.
As explained above, however, the interesting point is that nature can also work in
our favor here. Equation \eqref{Eq:NumEstimateAcc} only considers the most simple
case of an atom interacting with a single planar surface made from a homogeneous
and spatially local material.
In this perspective article, we have already pointed out that this is, at best,
only half of the complete picture.
Recent results based on the impact of long-time correlations on quantum friction
suggest that the acceleration experienced by the atom can be significantly enhanced
when
(i) the effects of spatial dispersion in the material are utilized,
(ii) geometric designs that suppress the net transfer of angular momentum are chosen,
and
(iii) the force's non-additive features are exploited to the fullest by, e.g.,
choosing cavities (parallel plates) or waveguides over simple planar interfaces.
By doing so, we venture to suggest that the quantum frictional acceleration to
be actually expected in experiments likely reads as
\begin{align}
  \tilde{a}\approx \eta\; a,
\end{align}
where $\eta$ denotes an enhancement factor relative to the result of the acceleration
as given in \eqref{Eq:NumEstimateAcc}.
A conservative estimate based on the discussion in Section \ref{Sec:III} gives
$\eta \gtrsim 2\times 10^4$ and additional room for further improvement seems
likely
~\cite{reiche21}.
We reiterate that $\eta$ is not achieved by optimizing a particular experimental
setting, but rather by grasping the deeper statistical nature of the interaction
to derive general arguments. We firmly believe that these results allow for modest
optimism that quantum friction might become experimentally accessible in the near
future.


\section{Concluding remarks}\label{Sec:conclusion}

Quantum friction describes the decelerating interaction between two or more objects
moving in vacuum relative to another. Similarly to the Casimir force and the van
der Waals / Casimir-Polder interaction, this drag is mediated by fluctuations of
the material-modified quantum electromagnetic vacuum and at zero temperature it is of a purely quantum mechanical origin.
Despite the fact that this phenomenon has continuously attracted attention
since the late 1970s, in recent years this attention has witnessed a quite significant
upsurge.
A number of investigations have appeared that reveal an underlying physics which
is considerably richer than one might have initially expected.
Even for one of the simplest and oldest issues in modern physics, i.e. the interaction
between a single atom and the quantized electromagnetic field, when the atom is moving
in the vicinity of a surface, it appears that well-established techniques and hard-earned
intuitions might be struggling in describing the behavior of the system.
The work of several authors has indeed uncovered a complex interplay of electromagnetism,
quantum physics, atomic physics, solid state physics and nonequilibrium statistical
physics which is intrinsic to quantum friction.
Rolling friction and a strong geometry-induced non-additivity are only some of the
consequences that originate from the merger of these fields of research.
It is due to such predictions which cannot be simply anticipated via equilibrium
physics, that experimental investigations of the force are strongly required and
eagerly awaited.
As we have pointed out in the present perspective, the current status
concerning this issue is reminiscent of the beginning of the 1930s to the end of 1940s
when the van der Waals interaction and the Casimir effect were predicted.
Both phenomena were unambiguously measured a few decades later and are nowadays
relevant in many nano-technological devices and applications.
The history of science is rich of cases where an interaction was deemed
to be too faint to be detected or even declared unmeasurable.
To recall one more example, the possibility of detection and even the existence
of gravitational waves was, for some time, famously doubted by the very person
that discovered them
~\cite{cervantes-cota16}.
Recently, he was proven wrong, at least in some cases, thereby opening new pathways to look at and to investigate the universe
~\cite{abbott16,yu20}.

As theorists, we would eventually and humbly like to offer the following perspective
on the strive for exploring and measuring quantum friction:
In the grand scheme of things, the experimental confirmation of the quantum frictional
force might not be the actual treasure. It is instead the access we will gain from this
to the underlying structure of quantum optical nonequilibrium fluctuations of the
material-modified vacuum. In simple terms, we would then have direct experimental
access to the very conceptional core of nonequilibrium quantum field theory. Testing
and probing its assumptions and consequences could be the first exciting step to new
frontiers, both theoretically and experimentally.


\begin{acknowledgments}
D.R. is indebted to Marty Oelschläger for countless inspiring discussions during a
joyful four-year-long professional endeavor.
We further thank Bettina Beverungen, Dan-Nha Hyunh, Christoph Egerland and Simon Herrmann for helpful exchanges as well as Bei-Lok Hu for input on the thermodynamic aspects of quantum friction.
We acknowledge support from the Deutsche Forschungsgemeinschaft (DFG, German Research Foundation) -- Project-ID 182087777 -- SFB 951.
F.I. acknowledges the hospitality and the financial support of the Erwin Schr\"odinger Institute.
\end{acknowledgments}

%


\end{document}